\documentclass[english]{revtex4-1}
\usepackage[latin1]{inputenc}
\usepackage{amsmath}
\usepackage{amsfonts}
\usepackage{amssymb}
\usepackage{graphicx}
\usepackage{hyperref}

\newcommand{\ld[1]}{\lambda_{#1}}
\newcommand{\lds}{\lambda_{345}}
\newcommand{\RD}{\Omega_{\text{DM}}h^2}
\newcommand{\Mh[1]}{M_{h_{#1}}}
\newcommand{\Mhc}{M_{h^{\pm}}}
\newcommand{\Mrho}{M_{\rho}}
\newcommand{\Mrhom}{M_{\rho}^{\text{MAX}}}

\begin{document}

\title{An i2HDM Strongly Coupled to a non-Abelian Vector Resonance}

\author{Felipe Rojas-Abatte} 
\email{astrofis.rojas@gmail.com}
\author{Maria Luisa Mora}
\email{maria.luisa.mora.u@gmail.com}
\author{Jose Urbina} 
\email{jose.urbina.avalos@gmail.com}
\author{Alfonso R. Zerwekh}
\email{alfonso.zerwekh@usm.cl}
\affiliation{Departamento de Fí­sica Universidad Técnica Federico Santa Marí­a
\\
and Centro Cientí­fico-Tecnológico de Valparaíso}

\begin{abstract}
We study the possibility of a Dark Matter candidate having its origin
in an extended Higgs sector which, at least partially, is related
to a new strongly interacting sector. More concretely, we consider
an i2HDM (\emph{i.e.} a Type-I Two Higgs Doublet Model supplemented with a $Z_{2}$ under which the non-standard scalar doublet is odd) based on the gauge group $SU(2)_{1}\times SU(2)_{2}\times U(1)_Y$. We assume that one of the scalar
doublets and the standard fermion transform non-trivially under
$SU(2)_{1}$ while the second doublet transforms under $SU(2)_{2}$.
Our main hypothesis is that standard sector is weakly coupled while
the gauge interactions associated to the second group is characterized
by a large coupling constant. We explore the consequences of this
construction for the phenomenology of the Dark Matter candidate and we show that the presence of the new vector resonance reduces the relic density saturation region, compared to the usual i2DHM, in the high Dark Matter mass range. In the collider side, we
argue that the mono-$Z$ production is the channel which offers the best chances to manifest the presence of the new vector field. We study the departures from the usual i2HDM predictions and show that the discovery of the heavy vector at the LHC is challenging even in the mono-$Z$ channel since the typical cross sections are of the order of $10^{-2}$ fb.
\end{abstract}

\maketitle

\section{Introduction}
The discovery of the Higgs boson \cite{Aad:2012tfa,Chatrchyan:2012xdj} crowned Standard Model (SM) with great success. However, the High Energy Physics community is unanimous to suspect that the SM is not a complete
description of the non-gravitational interactions. Three main open
questions justify this general conviction: a natural origin for
the electroweak scale, the origin of neutrino masses and the origin
of Dark Matter (DM). The first of these problems has motivated the
construction of many extensions of the SM. Some of them are based
on the elegant idea that the electroweak scale may be dynamically
produced in the context of a new strong interaction. Of course, this
proposal inevitably leads to the prediction a new composite sector.
On the other hand, although many observations point out to the existence
of DM, we have few clues about its nature. A very popular possibility
is that DM consists of neutral massive particles (with masses ranging
from some GeV's to some TeV's) with annihilation cross section of
the same order of magnitude than the cross sections obtained from
the weak interaction (the so called WIMP). One of the best known models that incorporate
this kind of DM candidate is a type-I 2HDM where one of the doublets
is odd under a new (and usually \emph{ad-hoc}) $Z_{2}$ symmetry. This model is
usually referred as the Inert Two Higgs Doublet Model or i2HDM \cite{Deshpande:1977rw, LopezHonorez:2006gr, Barbieri:2006dq}.
It is tempting to merge the ideas of an extended Higgs sector and
 compositeness, at least partially. Indeed already some authors have
explored the phenomenology of the 2HDM in the context of traditional dynamical electroweak symmetry breaking \cite{Luty:1990bg} and the so called
Composite Higgs Models where the scalar doublets arise as pseudo-Nambu-Goldstone
bosons \cite{Mrazek:2011iu, Bertuzzo:2012ya, DeCurtis:2016tsm, DeCurtis:2016gly, DiChiara:2016ybc, DeCurtis:2016scv}.
Additionally, for some particular models, it has been studied the
phenomenological consequences of a two Higgs doublet sector coupled
to composite vector resonances \cite{DiChiara:2016ybc, Zerwekh:2009yu}.
In this paper, we focus on a i2HDM where one of the scalar doublets 
( the one which is odd under the
$Z_{2}$ symmetry) is supposed to belong to a new strongly interacting
sector and is directly coupled to a vector resonance. This is in consonance with the very appealing idea of having a complex hidden sector with its own interactions and structure levels. We explore mainly
the consequences of the new heavy vector on the phenomenology of the
DM candidate. Additionally we argue that the best chance to observe a signature
of the new vector resonance at the LHC comes from the single production of a 
gauge boson plus missing transverse energy. To achieve our goals, we have organized our paper in the following way: in section \ref{Model} we describe our theoretical construction emphasizing the introduction of the new heavy vector. In section \ref{Constrains} we comment on the \emph{a priori} experimental and theoretical constrains whcih are relevant for our model. In section \ref{DMresults}, we describe our results for the phenomenology of the DM candidate while in section \ref{MonoZ} we focus on the mono-Z production at the LHC. Finally in section\ref{Conclusions} we state our conclusions.

\section{The Model}\label{Model}

Following the idea of Hidden Local Symmetry (HLS) \cite{PhysRevLett.54.1215},
we introduce the new vector resonance as the effective gauge fields
of a (hidden) gauge group which we call $SU(2)_{2}$. Consequently, our model is based on the local group $SU(2)_{1}\times SU(2)_{2}\times U(1)_{Y}$.
We assume the the first group is associated to the elementary or weak
interacting sector while the second group describes a composite or
strongly interacting sector. A fundamental hypothesis under our construction
is that standard left-handed fermions and one of the scalar doublets
($\phi_{1}$) transform under $SU(2)_{1}$ (and $U(1)_{Y}$) while
the second scalar doublet ($\phi_{2}$) transforms under $SU(2)_{2}$
(and the hypercharge group) as illustrated in Figure \ref{Moose}.
Additionally, we introduce a bi-doublet field which transforms as
$U_{1}\Sigma U_{2}^{\dagger}$ with $U_{1}$ and $U_{2}$ elements
of $SU(2)_{1}$ and $SU(2)_{2}$ respectively. With this ingredients,
and assuming that $\phi_{2}$ is odd under a new $Z_{2}$ symmetry,
the most general Lagrangian (with operators up to dimension 4) for
the gauge and scalar sector is:

\begin{figure}[ht]
\begin{center}
\includegraphics[scale=0.3]{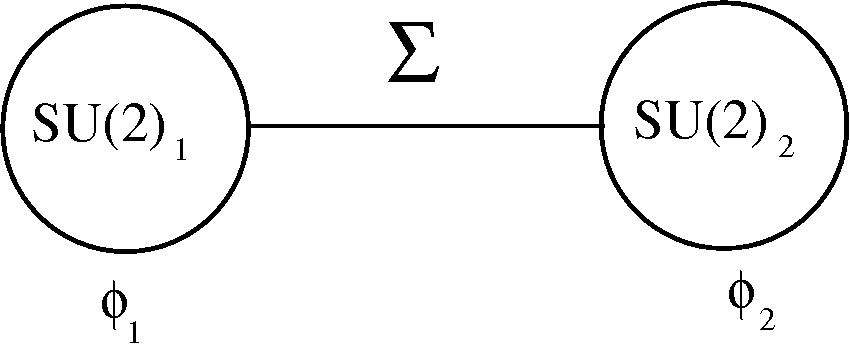}
\caption{Moose diagram representing the non-Abelian part of the group structure underlying our model. The $\Sigma$ link field is a bi-doublet while the scalars $\phi_1$ and $\phi_2$ are doublets of $SU(2)_{1}$ and $SU(2)_{2}$ respectively. } \label{Moose}
\end{center}
\end{figure}

\begin{eqnarray}
\mathcal{L} & = & -\frac{1}{2}Tr\left[F_{1\mu\nu}F_{1}^{\mu\nu}\right]-\frac{1}{2}Tr\left[F_{2\mu\nu}F_{2}^{\mu\nu}\right]+\frac{u^{2}}{2}Tr\left[\left(D_{\mu}\Sigma\right)^{\dagger}\left(D^{\mu}\Sigma\right)\right]\nonumber \\
 & + & \left(D_{\mu}\phi_{1}\right)^{\dagger}\left(D^{\mu}\phi_{1}\right)+\left(D_{\mu}\phi_{2}\right)^{\dagger}\left(D^{\mu}\phi_{2}\right)+m_{1}^{2}\left(\phi_{1}^{\dagger}\phi_{1}\right)+m_{2}^{2}\left(\phi_{2}^{\dagger}\phi_{2}\right)\nonumber \\
 & - & \lambda_{1}\left(\phi_{1}^{\dagger}\phi_{1}\right)^{2}-\lambda_{2}\left(\phi_{2}^{\dagger}\phi_{2}\right)^{2}-\lambda_{3}\left(\phi_{1}^{\dagger}\phi_{1}\right)\left(\phi_{2}^{\dagger}\phi_{2}\right)\nonumber \\
 & - & \lambda_{4}\left(\phi_{1}^{\dagger}\Sigma\phi_{2}\right)\left(\phi_{2}^{\dagger}\Sigma^{\dagger}\phi_{1}\right)-\frac{\lambda_{5}}{2}\left[\left(\phi_{1}^{\dagger}\Sigma\phi_{2}\right)^{2}+\left(\phi_{2}^{\dagger}\Sigma^{\dagger}\phi_{1}\right)^{2}\right]\label{eq:Lag1}
\end{eqnarray}

where

\[
D_{\mu}\Sigma=\partial_{\mu}\Sigma-ig_{1}A_{1\mu}\Sigma+ig_{2}\Sigma A_{2\mu}
\]

\[
D_{\mu}\phi_{j}=\partial_{\mu}\phi_{j}-ig_{j}A_{j\mu}\phi_{j}
\]
\noindent
and $u$ is an energy scale which characterize the new strong sector.

The $SU(2)_{1}\times SU(2)_{2}$ is spontaneously broken down to the
diagonal subgroup, which we identify with $SU(2)_{L}$, when the $\Sigma$
field acquires a v.e.v $\left\langle \Sigma\right\rangle =1$. In
this phase, Lagrangian (\ref{eq:Lag1}) becomes:

\begin{eqnarray}
\mathcal{L} & = & -\frac{1}{2}Tr\left[F_{1\mu\nu}F_{1}^{\mu\nu}\right]-\frac{1}{2}Tr\left[F_{2\mu\nu}F_{2}^{\mu\nu}\right]+\frac{u^{2}}{2}Tr\left[\left(g_{1}A_{1\mu}-g_{2}A_{2\mu}\right)\left(g_{1}A_{1}^{\mu}-g_{2}A_{2}^{\mu}\right)\right]\nonumber \\
 & + & \left(D_{\mu}\phi_{1}\right)^{\dagger}\left(D^{\mu}\phi_{1}\right)+\left(D_{\mu}\phi_{2}\right)^{\dagger}\left(D^{\mu}\phi_{2}\right)+m_{1}^{2}\left(\phi_{1}^{\dagger}\phi_{1}\right)+m_{2}^{2}\left(\phi_{2}^{\dagger}\phi_{2}\right)\nonumber \\
 & - & \lambda_{1}\left(\phi_{1}^{\dagger}\phi_{1}\right)^{2}-\lambda_{2}\left(\phi_{2}^{\dagger}\phi_{2}\right)^{2}-\lambda_{3}\left(\phi_{1}^{\dagger}\phi_{1}\right)\left(\phi_{2}^{\dagger}\phi_{2}\right)\nonumber \\
 & - & \lambda_{4}\left(\phi_{1}^{\dagger}\phi\right)\left(\phi_{2}^{\dagger}\phi_{1}\right)-\frac{\lambda_{5}}{2}\left[\left(\phi_{1}^{\dagger}\phi_{2}\right)^{2}+\left(\phi_{2}^{\dagger}\phi_{1}\right)^{2}\right]\label{eq:Lag2}
\end{eqnarray}
\noindent
and a mass mixing term appears in the gauge sector. On the other hand, the electroweak
symmetry breaking occurs, as in the SM, when $\phi_{1}$ gets a v.e.v: $\left\langle \phi_{1}\right\rangle =(0,v/\sqrt{2})^{T}$.
Notice that the $Z_{2}$ symmetry prevents $\text{\ensuremath{\phi}}_{2}$
from acquiring a v.e.v. This fact assures that $\phi_1$ is the SM Higgs doublet and forbid the appearance of any mass mixing term in the scalar sector. Finally, notice that, because of the same $Z_2$ symmetry, Yukawa terms can only be constructed  with $\phi_1$.

After these symmetry breaking processes, the following non-diagonal
mass matrices are generated for the neutral and charged vector bosons:

\[
M_{N}^{2}=\frac{v^{2}}{4}\left[\begin{array}{ccc}
(1+a^{2})g_{1}^{2} & -a^{2}g_{1}g_{2} & -g_{1}g_{y}\\
-a^{2}g_{1}g_{2} & a^{2}g_{2}^{2} & 0\\
-g_{1}g_{y} & 0 & g_{Y}^{2}
\end{array}\right]
\]

\[
M_{C}^{2}=\frac{v^{2}}{4}\left[\begin{array}{cc}
(1+a^{2})g_{1}^{2} & -a^{2}gg_{2}\\
-a^{2}g_{1}g_{2} & a^{2}g_{2}^{2}
\end{array}\right]
\]
\noindent
where $a=u/v$ and $g_{1}$, $g_{2}$ and $g_{Y}$ are the coupling
constants associated to $SU(2)_{1}$, $SU(2)_{2}$ and $U(1)_{Y}$.
When $M_{N}^{2}$ is diagonalized in the limit where $g_{2}\gg g_{1}$,
we obtain the following mass eigenstates for the neutral sector:

\begin{eqnarray*}
A_{\mu} & = & \frac{g_{Y}}{\sqrt{g_{1}^{2}+g_{Y}^{2}}}A_{1\mu}^{3}+\frac{g_{1}g_{Y}}{g_{2}\sqrt{g_{1}^{2}+g_{Y}^{2}}}A_{2\mu}^{3}+\frac{g_{1}}{\sqrt{g_{1}^{2}+g_{Y}^{2}}}B_{\mu}\\
Z_{\mu} & = & -\frac{g_{1}}{\sqrt{g_{1}^{2}+g_{Y}^{2}}}A_{1\mu}^{3}-\frac{g_{1}^{2}}{g_{2}\sqrt{g_{1}^{2}+g_{Y}^{2}}}A_{2\mu}^{3}+\frac{g_{Y}}{\sqrt{g_{1}^{2}+g_{Y}^{2}}}B_{\mu}\\
\rho_{\mu}^{0} & = & -\frac{g_{1}}{g_{2}}A_{1\mu}^{3}+A_{2\mu}^{3}.
\end{eqnarray*}
\noindent
where $\rho$ denotes the new heavy vector resonance. 

Similarly, the eigenstates of the charged sector (in the same limit)
are:

\begin{eqnarray*}
W_{\mu}^{\pm} & = & A_{1\mu}^{\pm}+\frac{g_{1}}{g_{2}}A_{2\mu}^{\pm}\\
\rho_{\mu}^{\pm} & = & -\frac{g_{1}}{g_{2}}A_{1\mu}^{\pm}+A_{2\mu}^{\pm}
\end{eqnarray*}
\noindent
where, as usual, $A_{n\mu}^{\pm}=\frac{1}{\sqrt{2}}\left(A_{n\mu}^{1}\mp A_{n\mu}^{2}\right)$.

In the same limit, the masses of the vector states can me expressed as:

\begin{eqnarray}
M_A&=&0 \qquad \mathrm{(exact)}\\
M_Z&\approx& \frac{v\sqrt{g_{1}^{2}+g_{y}^{2}}}{2}\left[1-\frac{1}{2}\frac{g_{1}^{4}}{g_{2}^{2}(g_{1}^{2}+g_{y}^{2})}\right]\\
M_{\rho^0}&\approx& \frac{a v g_2}{2}\left[1+\frac{g_1^2}{2g_2^2} \right]\\
M_{W}&\approx&\frac{v g_1}{2}\left[1-\frac{g_1^2}{2g_2^2} \right]\\
M_{\rho{\pm}}&\approx& \frac{a v g_2}{2}\left[1+\frac{g_1^2}{2g_2^2} \right]
\end{eqnarray}

Notice that to first order in $g_1/g_2$, we can write:

$$
\frac{g_1}{g_2}\approx a\frac{M_W}{M_{\rho}}
$$

The quantity $g_1/g_2$ is supposed to be small. This is the precise meaning of the assumption that the non-standard sector is strongly interacting. As we will explain below, 
in this work we consider values of $M_{\rho}$ in the 2-4 TeV range and $a=3,4,5$, obtaining $g_1/g_2 < 0.2 $.

In the scalar sector, the spectrum is straightforward since, as we already emphasized, no mass mixing term arise due to the $Z_2$ symmetry. Consequently, near the minimum of the potential, the scalar doublets can be parametrized as:

\begin{equation}
	\phi_{1}=\frac{1}{\sqrt{2}}\left(\begin{array}{c}
	0\\
	v+H
	\end{array}\right)\qquad\phi_{2}=\frac{1}{\sqrt{2}}\left(\begin{array}{c}
	\sqrt{2}h^{+}\\
	h_{1}+ih_{2}
	\end{array}\right)
\end{equation}
where $H$ is the SM-like Higgs boson and is identified with the observed 125 GeV scalar state. Notice that the $Z_2$ symmetry makes the lightest component of $\phi_2$ stable. As it is usually done, we assume that $h_1$ is the stable state and, consequently, the DM candidate.

Our model has seven free parameters: $u$, $g_{2}$, $m_{2}$ and
$\lambda_{i}$ with $i=2...5$ ($\lambda_1$ is fixed by the mass of the 125 GeV scalar observed at the LHC), however not all of them are equally
significant for our research. It is convenient, for phenomenological proposes to work with the following parameters:
\begin{equation}
\Mh[1], \quad \Mh[2], \quad \Mhc, \quad \Mrho, \quad \lds, \quad \ld[2], \quad a  \label{model-parameters}
\end{equation}
where $\Mh[1], \Mh[2], \Mhc$ are the physical masses of the new scalars, $\Mrho$ is the mass of the vector resonance and $\lds=\ld[3]+\ld[4]+\ld[5]$. Notice that $\lds$ plays a crucial roll controlling the interaction between the dark matter and the SM Higgs. According to this, we can rewrite the coupling constants as a function of the free parameters
\begin{eqnarray}
\nonumber &\displaystyle \ld[3] = \lds + 2\frac{\Mhc^2 - \Mh[1]^2}{v^2} \qquad \ld[4] = \frac{\Mh[1]^2 + \Mh[2]^2 - 2\Mhc^2}{v^2} \qquad \ld[5] = -\frac{\Mh[2]^2-\Mh[1]^2}{v^2}& \label{lambda_parameter} \\
& \displaystyle m_2^2 = \lds\frac{v^2}{2} - \Mh[1]^2  \qquad g_2 = \frac{2\Mrho}{va}& \label{eq:cond}
\end{eqnarray}

\section{Experimental and Theoretical Constrains}\label{Constrains}

The model parameter space can be constrained from theoretical restrictions coming from the analysis of the potential and experimental searches as well. In this section we mention all the restriction that we are account.
\begin{itemize}
\item \textbf{Vacuum stability:} In order to perform calculations around a minimum point without loose stability of the potential we need that there is no direction in field space along which the potential tends to minus infinity. This leads to the well-known conditions \cite{Branco:2011iw}
\begin{equation}
\ld[1] >0, \qquad \ld[2] >0, \qquad  2\sqrt{\ld[1]\ld[2]} + \ld[3] > 0, \qquad 2\sqrt{\ld[1]\ld[2]} + \ld[3] + \ld[4] + \ld[5] > 0  \label{rest:below_lim}
\end{equation}
\item \textbf{Neutral vacuum:} Another important requirement is that the vacuum must be electrically neutral. This can be guarantied if
\begin{equation}
\ld[5] < 0 \qquad \text{and} \qquad \ld[4]+\ld[5] < 0  \label{neutral-vacuum}
\end{equation}
The last condition (Eq.(\ref{neutral-vacuum})) assures us that $\Mh[1]$ is the lightest  particle which is odd under the $Z_2$ symmetry.
\item \textbf{Inert vacuum:} We need to consider the case where only the standard model field $\phi_1$ get a vacuum expectation value in order to avoid a mixing term between dark matter and the Higgs boson which will be catastrophic for abundance of relic density. According to reference \cite{Ginzburg:2010wa} the vacuum stability condition is satisfied provided that:
\begin{equation}
m_1^2 > 0 \qquad \text{and} \qquad m_2^2 < \sqrt{\frac{\ld[2]}{\ld[1]}}m_1^2
\end{equation}
In terms of our set of independent parameters, these conditions translate into:

 \begin{equation}
 \Mh[1]^2>\frac{v^2}{2}\left(\lds-2\sqrt{\ld[1]\ld[2]}\right)
 \end{equation}
 
 This is a very important constraint because it places an upper bound on $\lds$ for a given DM mass $\Mh[1]$.
\item \textbf{Perturbatibity:} All the quartic couplings of the potential must be limited by perturbatibity constraint, therefore 
\begin{equation}
|\ld[i]| \leq 8\pi
\end{equation} 
\item \textbf{Unitarity:} According to reference~\cite{Arhrib:2012ia} we can impose tree-level unitarity constraints if the eigenvectors of the scattering matrix elements between scalars and gauge bosons satisfy
\begin{equation}
|e_i|\leq 8\pi  \label{unitarity}
\end{equation}
where the parameters $e_i$ are defined as
\begin{eqnarray}
&& e_{1,2}= \ld[3] \pm \ld[4] \quad, \quad e_{3,4} = \ld[3] \pm \ld[5] \\
&& e_{5,6}= \ld[3] + 2\ld[4] \pm 3\ld[5] \quad, \quad e_{7,8} = -\ld[1]-\ld[2] \pm \sqrt{(\ld[1]-\ld[2])^2+\ld[4]^2} \\
&& e_{9,10} = -3\ld[1] - 3\ld[2] \pm \sqrt{9(\ld[1] -\ld[2])^2 + (2\ld[3]+\ld[4])^2}  \label{e9-10} \\
&& e_{11,12} = -\ld[1] - \ld[2] \pm \sqrt{(\ld[1] - \ld[2])^2 + \ld[5]^2}
\end{eqnarray}
\item \textbf{Electroweak precision Test:} In the i2HDM the electroweak radiative corrections are affected by the relation between the scalar masses~\cite{Barbieri:2006dq} alongside the Higgs mass and Z boson mass. The expressions for the S and T values are:
\begin{equation}
S = 
\frac{1}{72\pi}\frac{1}{(x_2^2-x_1^2)^3}
\left[ 
x_2^6 f_a(x_2) -x_1^6 f_a(x_1)
+ 9 x_2^2 x_1^2( x_2^2 f_b(x_2) - x_1^2 f_b(x_1)
\right]
\end{equation}
where $x_1=\frac{\Mh[1]}{\Mhc}, x_2=\frac{\Mh[2]}{\Mhc}, f_a(x) = -5+12\log(x), f_b(x)=3-4\log(x)$ and
\begin{equation}
T = \frac{1}{32\pi^2\alpha v^2}\left[F(\Mhc^2,\Mh[2]^2) + F(\Mhc^2,\Mh[1]^2) - F(\Mh[2]^2,\Mh[1]^2)\right]
\end{equation}
where the function $F(x,y)$ is defined by
\begin{equation*}
F(x,y) = 
\begin{cases}
\frac{x+y}{2}-\frac{xy}{x-y}\log{\left(\frac{x}{y}\right)}, & x\neq y\\
0, & x = y
\end{cases}
\end{equation*}
Written in this form, according to Ref~\cite{Belyaev:2016lok}, the contribution to $S$ and $T$ shows explicitly that we cannot distinguish the CP properties of $h_1$ and $h_2$. With $U$ fixed to be zero, the central value of $S$ and $T$, assuming a SM Higgs boson mass of $m_h$ = 125 GeV, are given by~\cite{Baak:2014ora}
\begin{equation}
S = 0.06 \pm 0.09 ,\qquad T = 0.1 \pm 0.07  \label{EWPT}
\end{equation}
with the correlation coefficient +0.91.
\item \textbf{LHC constrains on vector resonances:} In general, vector resonances  may produce detectable signals at colliders through channels like dijet production, dilepton production, the associate production of a Higgs boson and a gauge boson, and the production of two gauge bosons. Also the Higgs decay rate into two photons (which is loop process) and the oblique parameter $S$, $T$ may receive sensible corrections from heavy charged fields. However, in our case the new vector resonance couples to the SM fields only through mixing terms which are suppressed by factors $g_1/g_2$. Moreover, previous studies suggest that the experimental constrains are largely satisfied if the new resonance is heavier than 2.4 TeV \cite{Castillo-Felisola:2013jua, Carcamo-Hernandez:2013ypa, Gintner:2017cfg}. 
\begin{figure}[ht]
\begin{center}
	\includegraphics[scale=0.22]{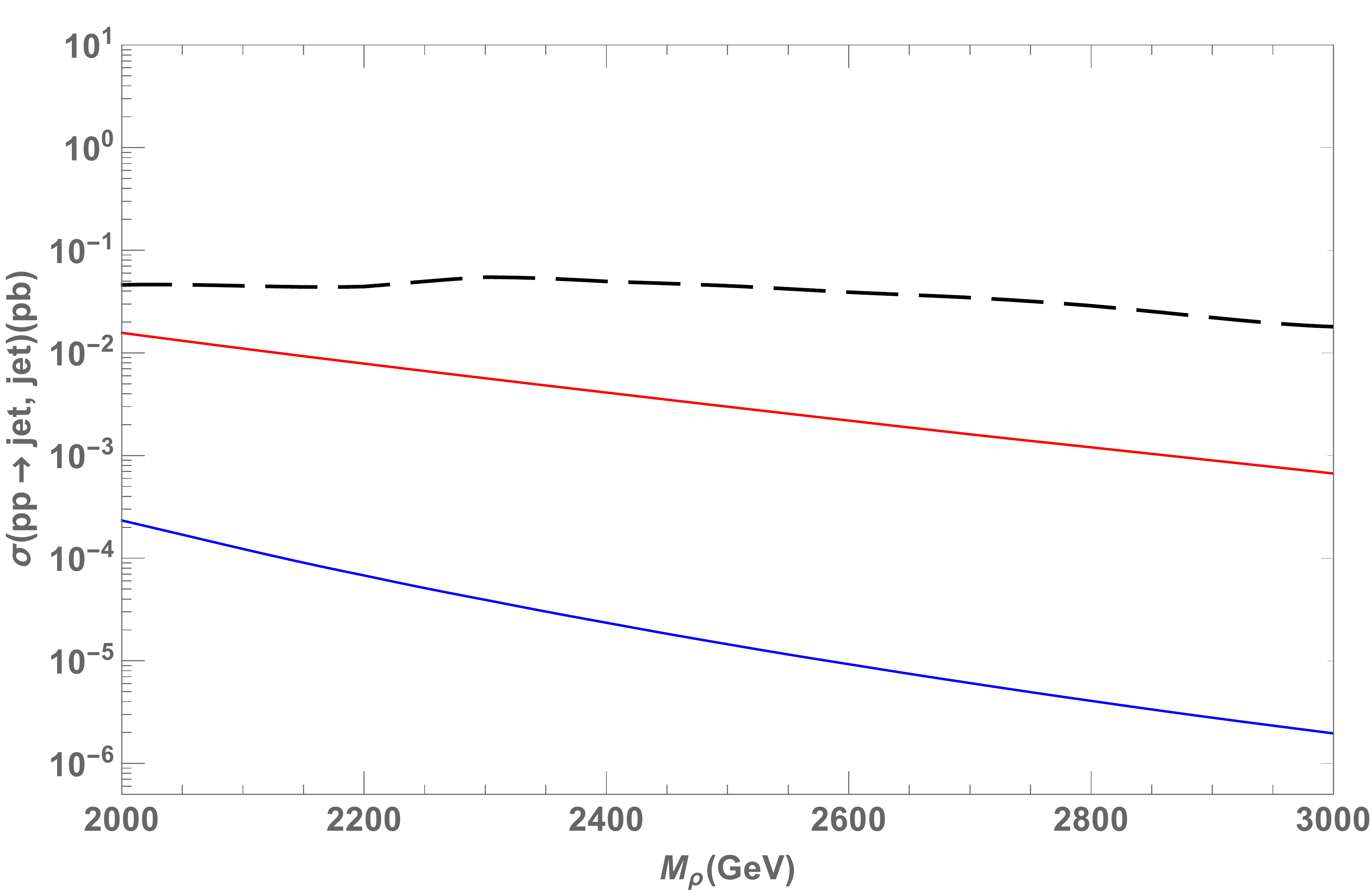}
	\caption{$\sigma(pp \rightarrow \rho_{\mu} \rightarrow j j)$ computed in the kinamatic region where the $\rho_{\mu}$ decay channels into a pair of non-standard scalars are open (lower solid line) or closed (higher solid line). $a=3$}\label{fig:dijet}
\end{center}
\end{figure}
As a matter of example, we compare the cross section predicted by our model for the process $pp \rightarrow \rho_{\mu} \rightarrow j j$ with the upper limits set by ATLAS  for dijet resonances \cite{ATLAS:2016lvi}, as shown in Figure \ref{fig:dijet}. Our calculations are performed in two different kinematic regimes depending on whether the $\rho_{\mu}$ decay channels into a pair of non-standard scalars are open or not. When these channels are open they dominate over the decay into SM particles since the interaction in the former case is proportional to $g_2$ while in the latter case is suppressed by a factor $g_1/g_2$. This makes the resonant dijet production quit unprovable as shown by the lowest continuous line in Figure \ref{fig:dijet}. The upper continuous line, on the other hand, shows the predicted cross section when the vector resonance is not able to decay into non-standard scalars. Notice that in the appropriate kinematic regime, values of $\Mrho < 2.4$ TeV are allowed.

\item \textbf{LHC limits from Higgs di-photon decay:} The decay rate of the Higgs bosons into two photons does not constrain very much the mass of the vector resonance either because the Higgs boson couples to $\rho_{\mu}$ only as a result of the mixing between $A_{1\mu}$ and $A_{2\mu}$ and, consequently, the $H\rho_{\mu}^{+}\rho_{\nu}^{-}$ vertex is suppressed by a factor $(g_1/g_2)^2$. However, the interaction vertex $Hh_{\mu}^{+}h_{\nu}^{-}$ is governed by the $\ld[3]$ quartic coupling which can be constrained through loop calculations. We can use the limit coming from ATLAS-CMS Higgs data analysis~\cite{ATLAS-and-CMS} to set a restriction on $\ld[3]$ using the experimental value: 
\begin{equation}
\frac{Br^{BSM}(H\rightarrow\gamma\gamma)}{Br^{SM}(H\rightarrow\gamma\gamma)} = \mu^{\gamma\gamma} = 1.16^{+40}_{-36}  \label{higgs-aa}
\end{equation}
\item \textbf{Invisible Higgs-decay:} Interactions among Higgs boson and the new sector (inert scalars and vector resonance) are allowed in this model, therefore the possibility of new invisible decay channels are open. Those channels could lead to deviations of Higgs boson decay width from the SM value. Using results that comes from ATLAS~\cite{Aad:2015txa} at 95\% CL we can restrict the invisible Higgs decay to be less than
\begin{equation}
Br(H\rightarrow \text{invisible}) < 28\% \label{Br-inv}
\end{equation}
which is also compatible with the CMS result~\cite{CMS-PAS-HIG-15-012}. 
\item \textbf{LEP limits on inert scalars:} In order to not affect the precise measurements of W and Z widths we need to impose restrictions to the mass of the inert scalars demanding that $\Gamma(W^{\pm}\rightarrow h_1h^{\pm})$, $\Gamma(W^{\pm}\rightarrow h_2h^{\pm})$, $\Gamma(Z\rightarrow h_1h_2)$ and $\Gamma(Z\rightarrow h^+h^-)$ channels are kinematically closed.
This leads to the following constraints:
\begin{eqnarray}
\nonumber &\Mh[1] + \Mhc > M_{W^{\pm}} \qquad \Mh[2] + \Mhc > M_{W^{\pm}}& \\
&\Mh[1] + \Mh[2] > M_{Z} \qquad 2\Mhc > M_{Z}&  \label{LEP}
\end{eqnarray}
\item \textbf{Relic Density limits:} We analyze the abundance of dark matter using \texttt{micrOMEGAs}~\cite{Belanger:2013oya, Belanger:2006is, Belanger:2010gh} package. This program solves the Boltzmann equation numerically, using \texttt{CalcHEP}~\cite{Belyaev:2012qa} to calculate all of the relevant cross sections. The program consider the case when $M_{h_1}<M_W,M_Z$ taking into account the annihilation into 3-body final state from $VV^*$ or 4-body final state from $V^*V^*$ ($V=W^{\pm},Z$). Co-annihilation effects are taken into account as well. We require that our predictions for the relic density be in agreement with the PLANCK \cite{Ade:2013zuv, Planck:2015xua} measurement:
\begin{equation}
\Omega_{\text{DM}}^{\text{Planck}}h^2 = 0.1184 \pm 0.0012  \label{PLANCK-lim}
\end{equation}
\item \textbf{Direct Detection limits:} Using the first dark matter results coming from XENON1T~\cite{Aprile:2017iyp} with 34.2 live days of data acquired between November 2016 and January 2017 we have evaluated the spin-independent cross section of DM scattering off the proton, $\sigma_{SI}$, also using \texttt{micrOMEGAs}.
\end{itemize}

\section{Dark Matter Phenomenology}\label{DMresults}
As we explained above, our model has a 7-dimensional parameter space, however we can have a good phenomenological overview of the model focusing only on 3  specific parameters ($\lds$, $\Mh[1]$, $\Mrho$) and fixing all the other ones to which the phenomenological observables have poor sensibility. For instance, the dark matter candidates and the SM fields only interact through the Higgs boson, the electroweak gauge bosons and the new heavy vector; but, since the interaction with the standard gauge bosons is governed by the electroweak gauge couplings which are fixed, the only relevant free parameter is $\lds$, the dark matter mass itself ($\Mh[1]$) and $\Mrho$. 
\begin{figure}[htb]
\centering
{\includegraphics[width=0.48\textwidth]{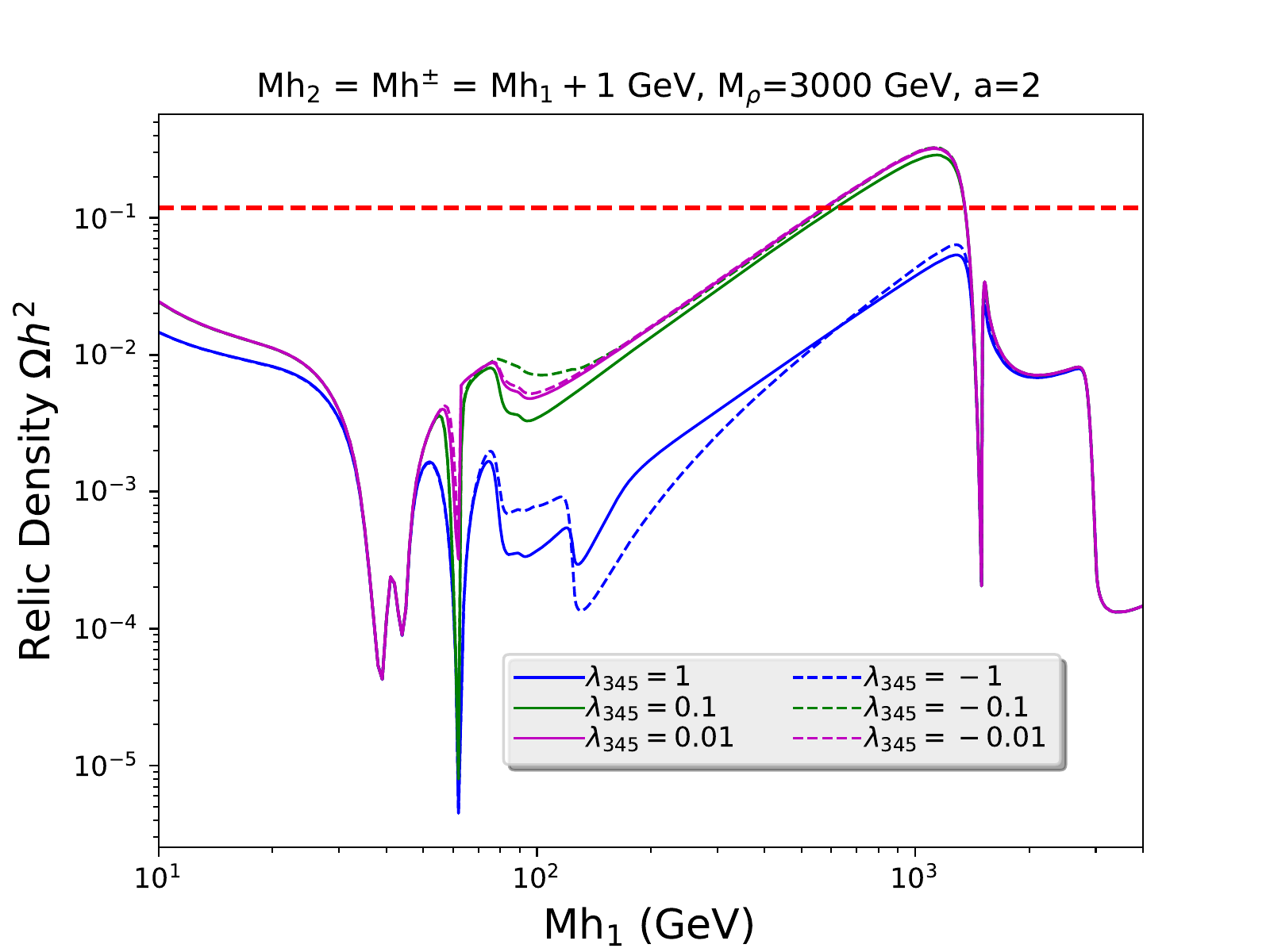}}%
{\includegraphics[width=0.48\textwidth]{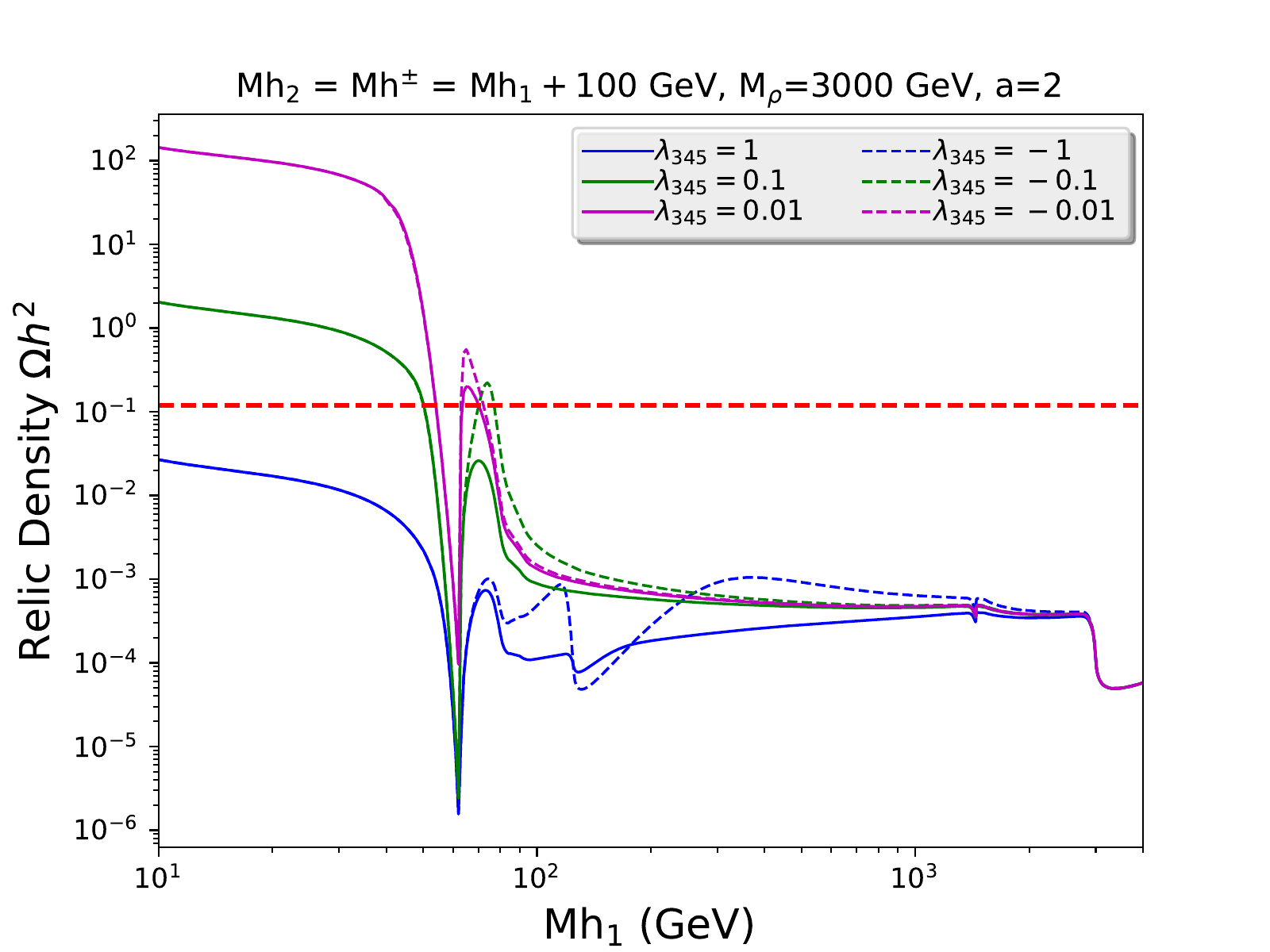}}%
\vskip -0.5cm\hspace*{-3cm}(a)\hspace*{0.48\textwidth}(b)
\caption{Relic density $\RD$, as a function of $\Mh[1]$ for different values of $\lds$ in a quasi-degenerate scenario (a) where $\Mh[2]=\Mhc = \Mh[1] + 1$ and a no-degenerate scenario (b) where $\Mh[2]=\Mhc = \Mh[1] + 100$. In both cases we fix the values of $\Mrho = 3000$ GeV, $a=2$ and $\ld[2]=1$. The horizontal red line corresponds to the relic density measurements PLANCK limits.} \label{fig:RD-mh1-lds}
\end{figure}

In Figure~\ref{fig:RD-mh1-lds} we show a 2-dimensional section of the  parameter space where we have the dark matter relic density as a function of $\Mh[1]$ for several values of $\lds$. For simplicity, in this analysis we always take $\Mh[2]=\Mhc$. With this assumption an important kinematic parameter is $\Delta M\equiv \Mh[2]-\Mh[1]$. Now, two qualitatively different scenarios can be distinguished: a quasi-degenerate case where $\Delta M = 1$ GeV and a non-degenerate case $\Delta M=100$ GeV. In both we considerer $\Mrho = 3000$ GeV, $a=2$ and $\ld[2]=1$. We can notice that for $10$ GeV $\leq \Mh[1] \ll \Mrho/2$ GeV (which we will refer as the low mass region) the model reproduces the same pattern of relic density predicted by the usual the i2HDM, as expected.  It is only when $\Mh[1]$ approaches to $\Mrho/2$ that the effect of the vector resonance $\rho$ becomes important.

In the reference \cite{Belyaev:2016lok} there is a detailed phenomenological explanation of what happens in the low mass region, so we will just briefly comment on it. Here, we can distinguish two different asymptotic behaviors: the first one for $10$ GeV $< \Mh[1] < 50$ GeV and the second one ($\Mh[1] > 200$) GeV. 

In Figure \ref{fig:RD-mh1-lds}a), which shows the quasi-degenerate case, we can see that  below 62.5 GeV (\textit{i.e} half of the Higgs boson mass) the co-annihilation effects between the inert scalars become important because of the appearance of new annihilation channels, pushing the DM Relic density under the experimental PLANCK limit. On the other hand, in the non-degenerate case (when $\Delta M = 100$ GeV), as seen in Figure \ref{fig:RD-mh1-lds}b), co-annihilation is suppressed generating an enhancement of the $\RD$ becoming even 3 orders of magnitude above the PLANCK limit for small values of $\lds$ ($\sim 0.01$). 

Now, in the second  case (\textit{i.e} for $\Mh[1] > 200$ GeV), when $\Delta M = 1$ GeV the quartic coupling becomes small enough to produce a significant suppression of the Dark Matter annihilation into longitudinal polarized gauge bosons. This effect increases the relic density which is capable of reaching the PLANCK limit even considering the effects of co-annihilation. On the other hand, for the non-degenerate case, as seen in Figure \ref{fig:RD-mh1-lds}b), the value $\Delta M$ is large and the average annihilation cross sections of the processes $h_1h_1\rightarrow W_LW_L$ and $h_1h_1\rightarrow Z_LZ_L$ are increased, making the abundance of relic density too low to reach the saturation limit. This generates the flat asymptotic behavior for large  values of $\Mh[1]$.

When $\RD$ reaches the PLANCK limit in the high mass region, but now considering the $\Delta M=1$ GeV case, the annihilation average cross section through the vector resonance starts to be important as the value of $\Mh[1]$ increases. At $\Mh[1] = \Mrho/2$ GeV the value of the relic density distribution decreases dramatically due to co-annihilation of $h_1$ and $h_2$ into an on-shell $\rho$ vector. The wide deep around 3000 GeV (see Figure \ref{fig:RD-mh1-lds}a)) corresponds to the opening of annihilation channels $h^+ h^-\rightarrow \rho \rightarrow \rho^+ \rho^-$, $h_1 h_1 \rightarrow \rho^+ \rho^-$ and $h_2 h_2 \rightarrow \rho^+ \rho^-$.  In the case where $\Delta M=100$ GeV, the main annihilation processes are $h_1h_1\rightarrow W^+W^-$ and $h_1h_1\rightarrow ZZ$, although there is a small contribution ($\sim 4\%$) of the process $h_1h^+ \rightarrow \rho^+ H$ via s-channel $\rho$ boson interchange which generate the small negative peak at $\Mh[1]=\Mrho/2$. Finally, in this case, the last deep at $\Mh[1]=3000$ GeV is produced through the opening of the annihilation channels $h_1h_1 \rightarrow \rho^+ \rho^-$ and $h_1h_1 \rightarrow \rho^0 \rho^0$ . 


In order to have a complete visualization of how the vector resonance affects the i2HDM, we performed a random scan over the 7-dimensional parameter space considering all the experimental and theoretical constraints mentioned in section \ref{Constrains}. In our analysis, we exclude all the points in the parameter space where over-abundance take place because they are considered non-physical. However, we keep the regions of points which produce under-abundance since it only means that an additional source of DM is needed. Consequently, we used a re-scaled Direct Detection cross section $\hat{\sigma}_{\text{SI}} = \left(\Omega_{\text{DM}}/\Omega_{\text{PLANCK}}\right) \times \sigma_{\text{SI}}$ which allows us to take into account the case when $h_1$ contribute only partially to the total amount of DM. The range of the scan for each free parameter is summarized en Table \ref{tab_param}. 

\begin{table}[ht]
	\caption{Range of the parameter space}
	\begin{center}
		\begin{tabular}{c|c|c}
			\hline
			\hline
			\textbf{Parameter}      & \textbf{min value} &  \textbf{max value} \\ \hline \hline 
			$\Mh[1]$ [GeV]   &    480	&     4500  \\ 
			$\Mh[2]$ [GeV]   &    480	&     4500  \\ 
			$\Mhc$ [GeV]&   480	&     4500  \\ 
			$\Mrho$ [GeV]&   2500	&     4500  \\ 
			$\lds$            &  -5    &     5  \\ 
			$\ld[2]$          &  0    	&     5  \\ 
			$a$		  &  3		&     5 \\ \hline \hline
		\end{tabular} 
	\end{center}
	\label{tab_param}
\end{table}

As it was previously explained, our model reproduces the same pattern of $\RD$ as the i2HDM for $\Mh[1]\ll\Mrho/2$ because the interaction between the SM particles and the vector resonance $(\rho_{\mu})$ is suppressed by the factor $(g_1/g_2)$. Therefore we will focus on the high mass region where the interaction with the vector resonance is more sensitive.

\begin{figure}[htb]
\centering
{\includegraphics[width=0.5\textwidth]{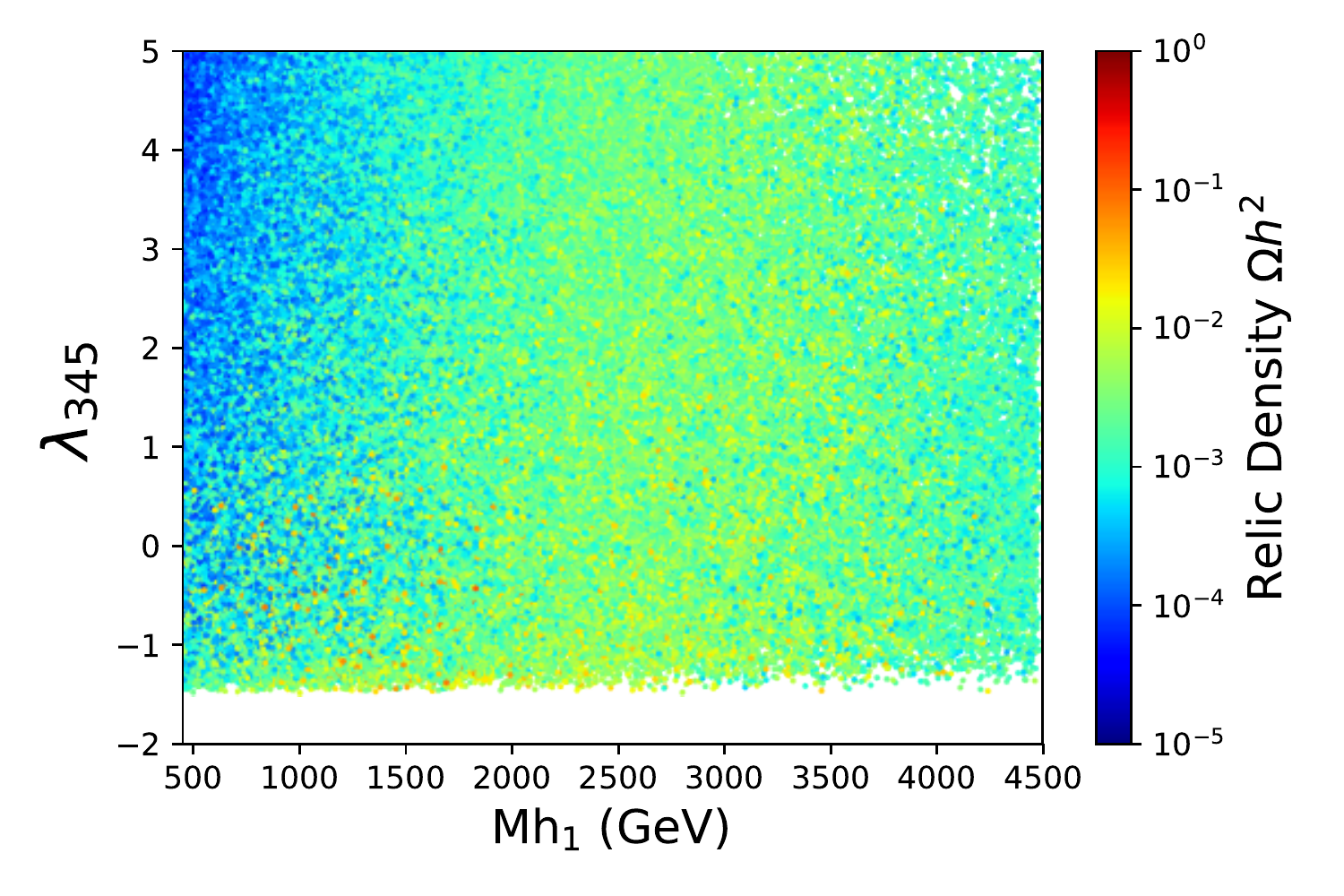}}%
{\includegraphics[width=0.5\textwidth]{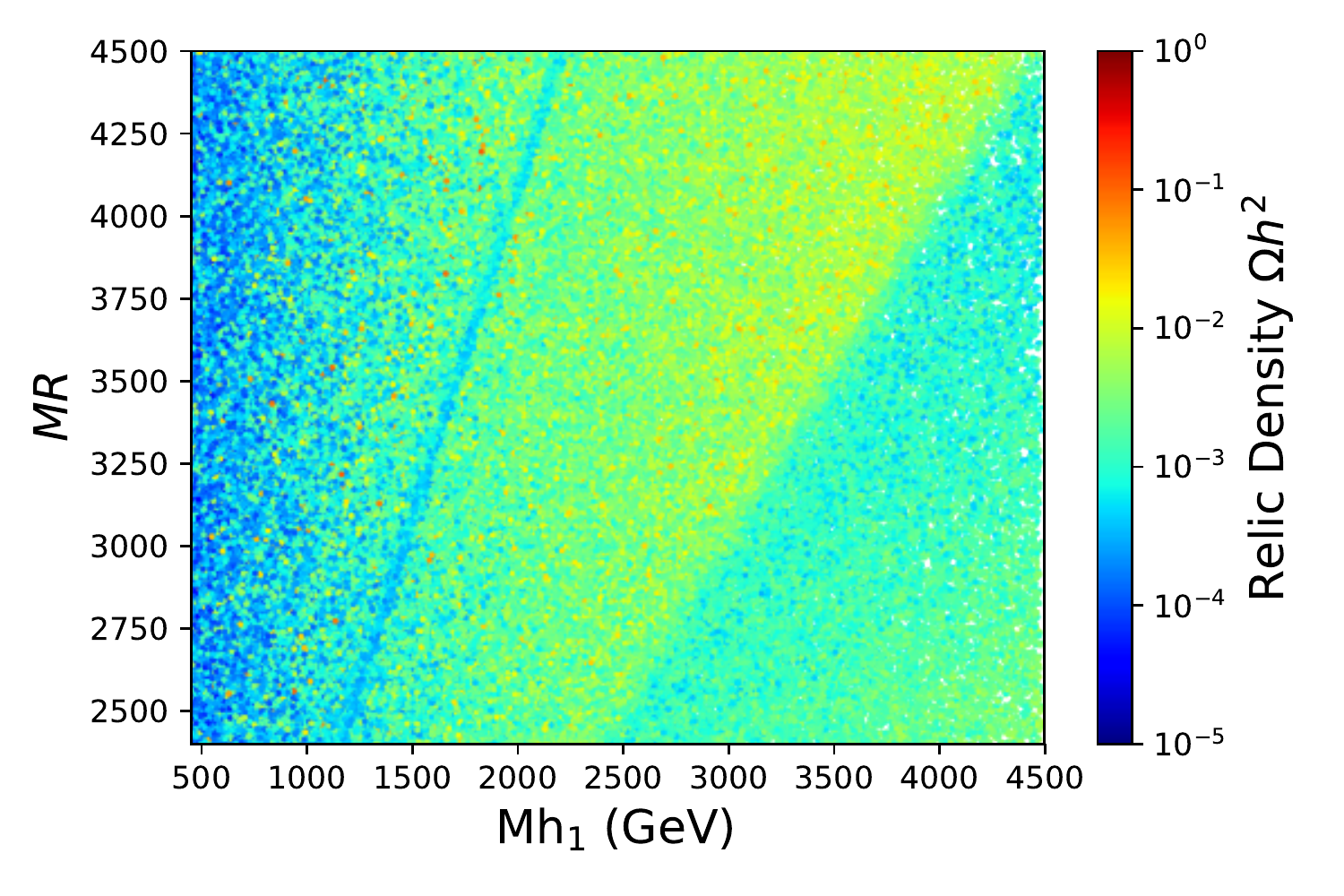}}%
\vskip -0.5cm\hspace*{-3cm}(a)\hspace*{0.48\textwidth}(b)
\caption{2D projections of the 7D random scan of the model parameter space restricted to (450 GeV, 4500 GeV) for $\Mh[1]$, (2500 GeV, 4500 GeV) for $\Mrho$ and (-2,5) for $\lds$ considering all constraints except under-abundance of DM. } \label{fig:high-mass-param}
\end{figure}

In Figure \ref{fig:high-mass-param}, we show projections in 2-dimensional planes of the scan as a color map of DM relic density where we show the planes ($\Mh[1],\lds$) and ($\Mh[1],\Mrho$). In Figure \ref{fig:high-mass-param}a), we can see the effect of the vacuum stability constraint on  $\lds$, making it to satisfiy the bound $\lds \gtrsim -1.47$.

It is easy to recognize the DM annihilation into an on-shell vector resonance $(h_1 h_2 \rightarrow \rho)$ at $\Mh[1]\approx \Mrho/2$ GeV through the substantial DM relic density decrease in a narrow sector represented by the diagonal blue pattern in Figure \ref{fig:high-mass-param}b).  

\begin{figure}[htb]
\centering
{\includegraphics[width=0.5\textwidth]{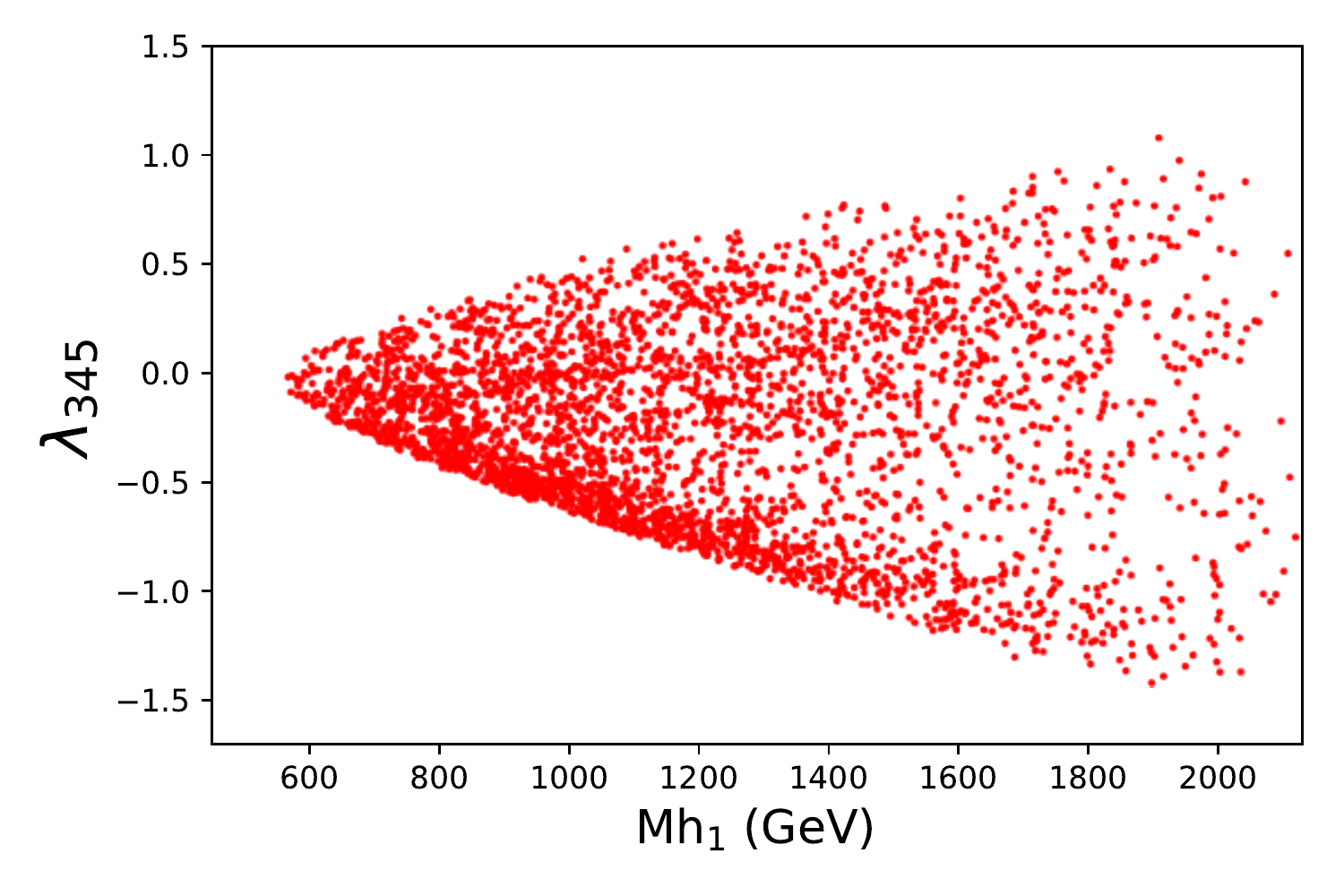}}%
{\includegraphics[width=0.5\textwidth]{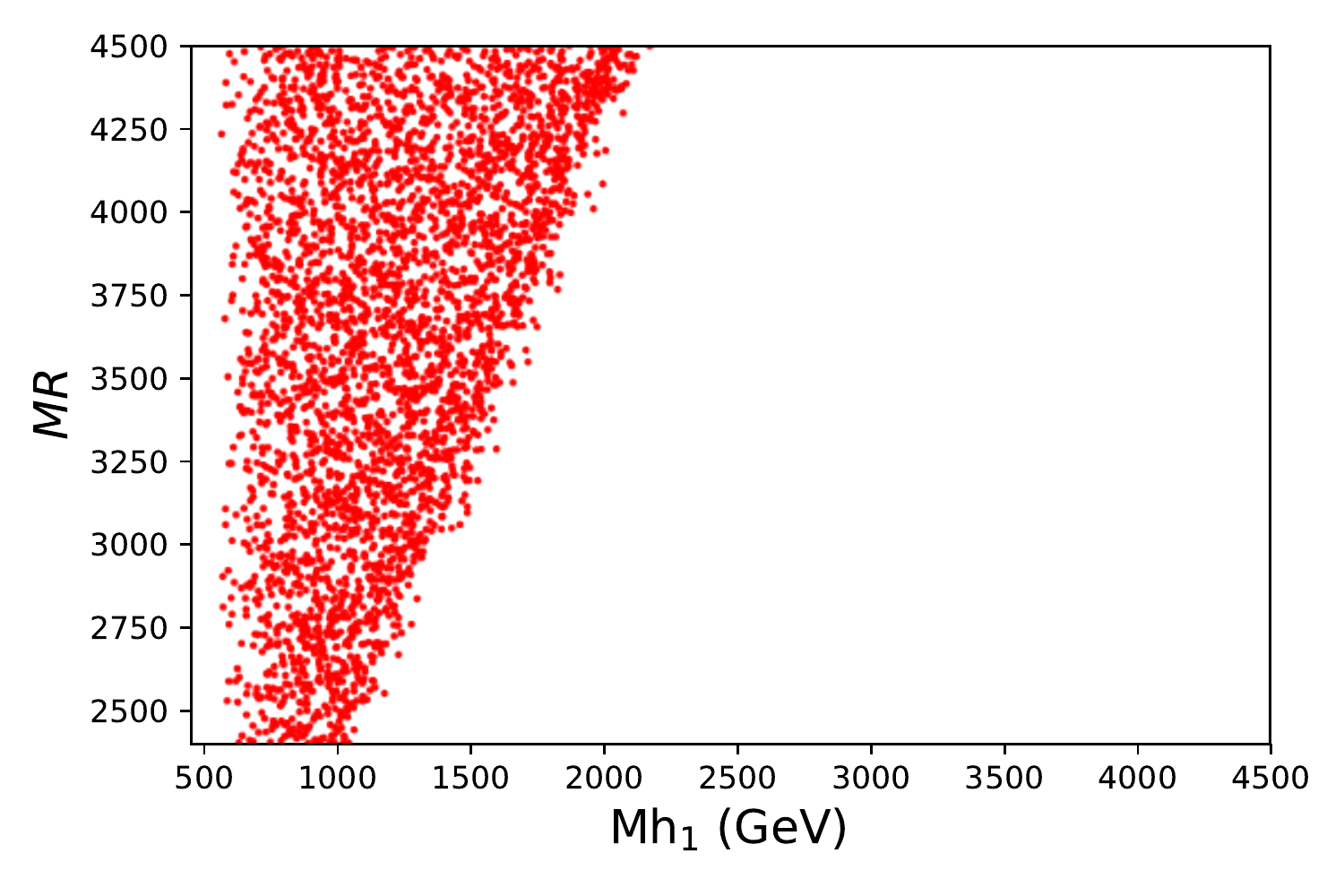}}%
\vskip -0.5cm\hspace*{-3cm}(a)\hspace*{0.48\textwidth}(b)
\caption{2D projections of the 7D random scan of the model parameter space restricted to (450 GeV, 4500 GeV) for $\Mh[1]$, (2500 GeV, 4500 GeV) for $\Mrho$ and (-2,5) for $\lds$ considering all constraints plus the lower PLANCK limit.} \label{fig:saturation}
\end{figure}

It is important to stress that $\Mrho/2$ establishes a border in the parameter space for the saturation of relic density. For $\Mh[1] > \Mrho$ , the annihilation cross section becomes more intense and the abundance of relic density decreases below the experimental PLANCK limit. This border is clearly seen in Figure \ref{fig:saturation}b) where we present the parameter space which at the same time reproduces the value of $\RD$ observed by PLANCK and is consistent with all the experimental constrains. In other words, the interactions due to the new vector resonance reduce the saturation region in the high mass zone compared to i2HDM because when the DM reaches the limit $\Mh[1] \approx \Mrho$ the channels $h_1 h_2 \rightarrow \rho^+\rho^-$ and $h_1 h_2 \rightarrow \rho^0\rho^0$  become open causing the abundance of DM to fall down by at least one order of magnitude, as we can clearly see from Figure \ref{fig:high-mass-param}b).

As we stressed before, in the high mass zone it is possible to reach the saturation limit of the relic density due to the high level of degeneracy of the three inert scalars, which turns out to be no more than a few GeV. This mass split is closely related to the quartic coupling of the potential. A small difference of mass implies  small values of the $\lambda$ parameters which translates into a low average annihilation cross section of the dark matter into longitudinal polarized gauge bosons, generating an enhancement in the abundance of relic density. This can be seen in Figures \ref{fig:RD-mh1-lds}a) and \ref{fig:saturation})a) where  $\lds$ can reach higher values as $\Mh[1]$ increases. This effect is maintained until the threshold is reached at $\Mh[1] = 2250 = \Mrhom/2$ GeV, where  $\Mrhom = 4500$ GeV is the maximum value of $\Mrho$ used in our parameter space.

\section{Predictions for the LHC: Mono-Z production}\label{MonoZ}

\begin{figure}[htb]
	\centering
	\includegraphics[scale=0.5]{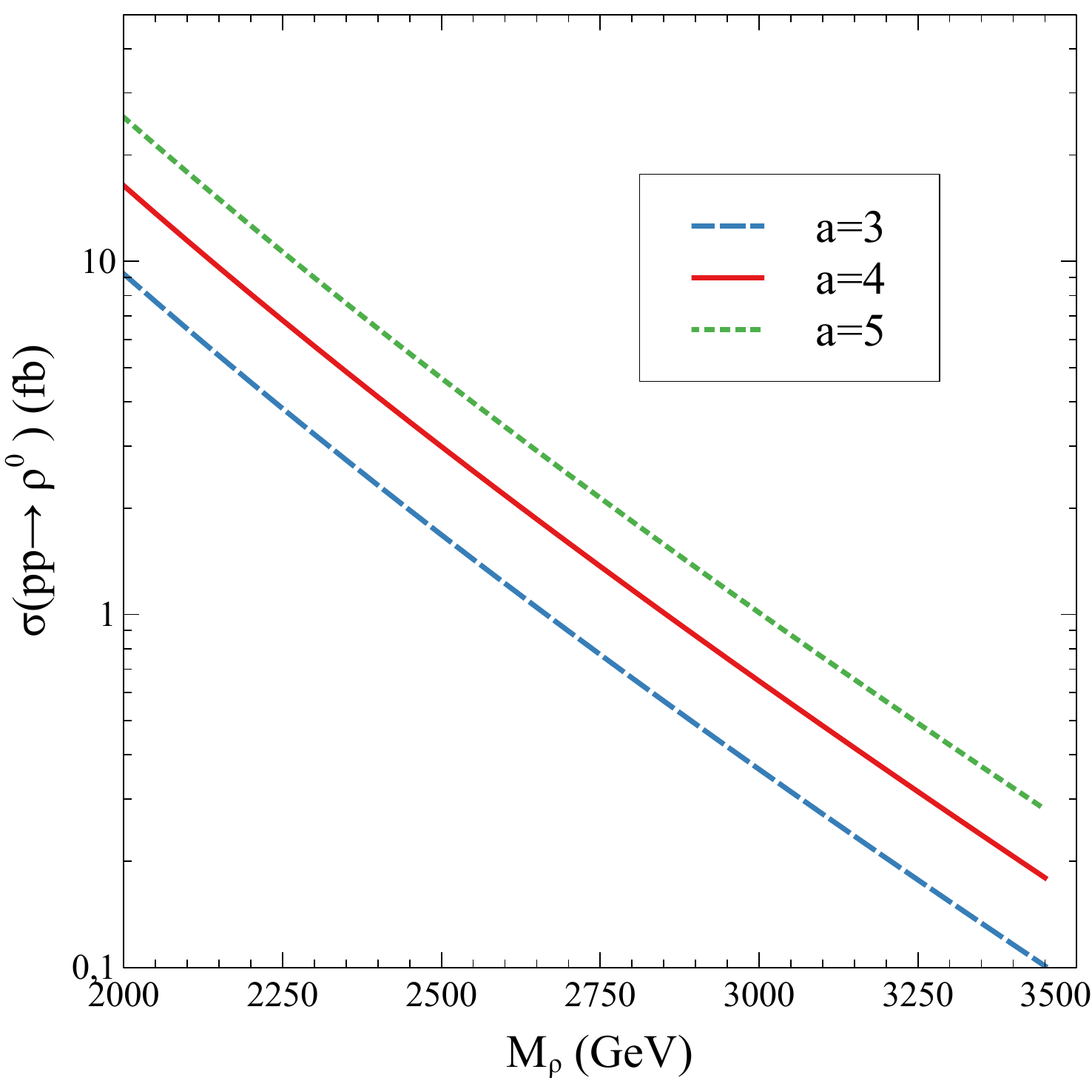}
	\caption{$\sigma(pp \rightarrow \rho^0)$ vs. $M\rho$ at the LHC for  $\sqrt{s}=13$ TeV }
	\label{fig:ProducionTotalRho}
\end{figure}

At the LHC, the new vector resonance is mainly produced by quark annihilation. In consequence, the total production cross section $\sigma(pp \rightarrow \rho)$ is proportional to $(g_1/g_2)^2 \approx a^2 M_W^2/M^2_{\rho} $. In Figure \ref{fig:ProducionTotalRho} we show our predictions for $\sigma(pp \rightarrow \rho^0)$ at the LHC with $\sqrt{s}=13$ TeV. The tiny cross sections indicate that it is a very challenging task to discover the new heavy vector at the LHC specially when we consider only standard particles in the final states, since the interaction of the heavy vector with particles of the SM is suppressed by factors $(g_1/g_2)$.

\begin{figure}	
	\includegraphics[scale=0.43]{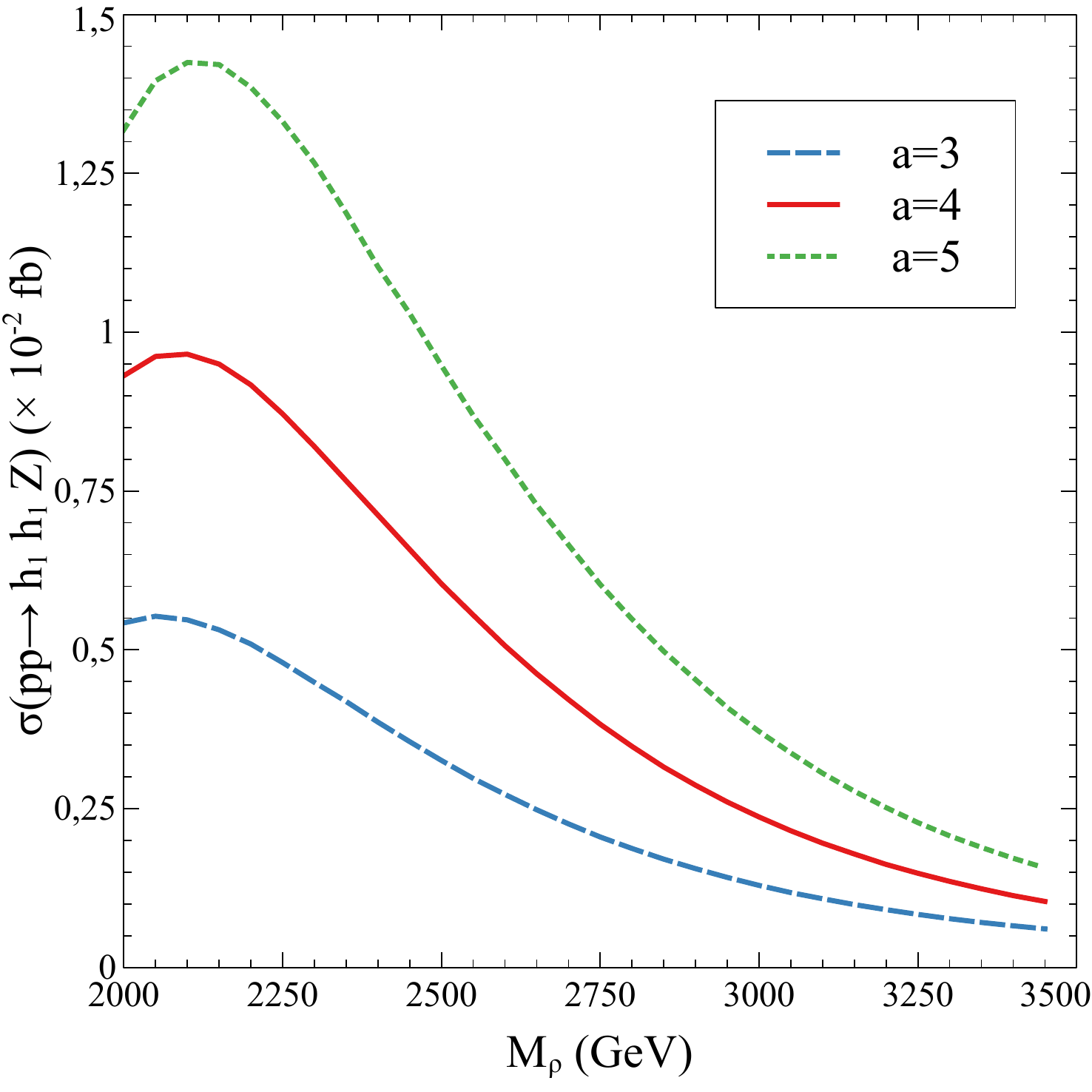}
	\includegraphics[scale=0.51]{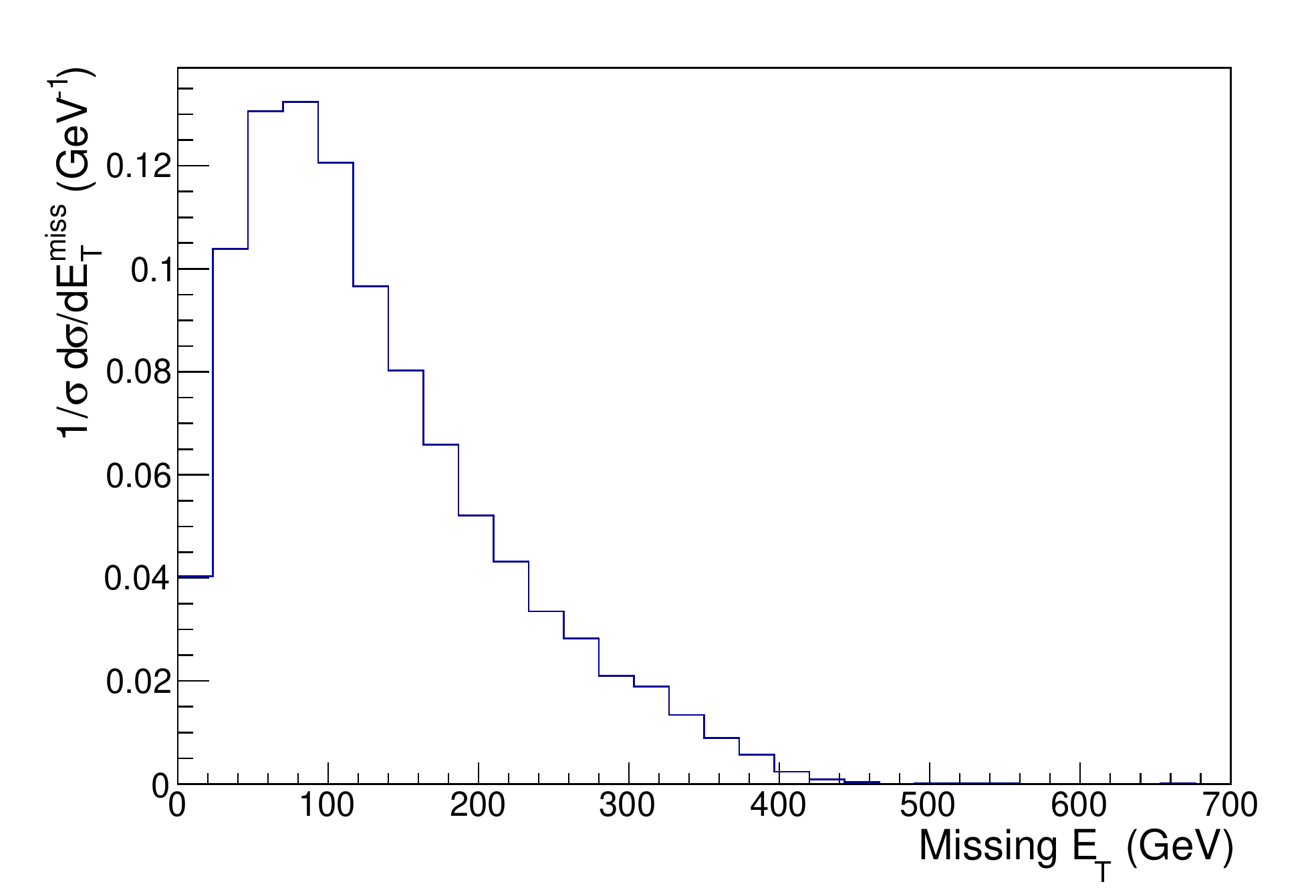}
	\caption{\textbf{Left}: $\sigma(pp\rightarrow h_1 h_1 Z)$ at the LHC for $\sqrt{s}=13$ TeV, $a=3$ (dashed), $a=4$ (continuous),$a=5$ (dotted). We use $M_{h1}=800$ GeV, $M_{h2}=810$ GeV, $\ld[345]=-0.1$ and $\ld[2]=2.0$. \textbf{Right}: Normalized missing $E_T$ distribution  }\label{fig:monoz}
\end{figure}

However, we can expect to have a better chance of getting observable signals if we consider final states containing the new scalar alongside some standard particle. A promising process is  $pp\rightarrow h_1 h_1 V$ (with $V=Z$ or $W^{\pm}$) . In this process the scalars are not detected but they produce a significant amount of missing transverse energy, as shown in Figure \ref{fig:monoz} (right). Hereafter, we focus on the mono-Z production. Figure \ref{fig:monoz} (left) shows the predicted cross section for the process $pp\rightarrow \rho \rightarrow h_1 h_1 Z$ computed for three values of the $a$ parameter ($a=3,4,5$) while other relevant parameters were took as $M_{h1}=800$ GeV, $M_{h2}=M_{h^{\pm}}=810$ GeV, $\ld[345]=-0.1$ and $\ld[2]=2.0$. As we see, for $M_{\rho}$ between 2 and 3.5 TeV, the cross section lies in the range of $(0.05-1.5)\times10^{-2}$ fb.

\begin{figure}
	\includegraphics[width=0.45\textwidth]{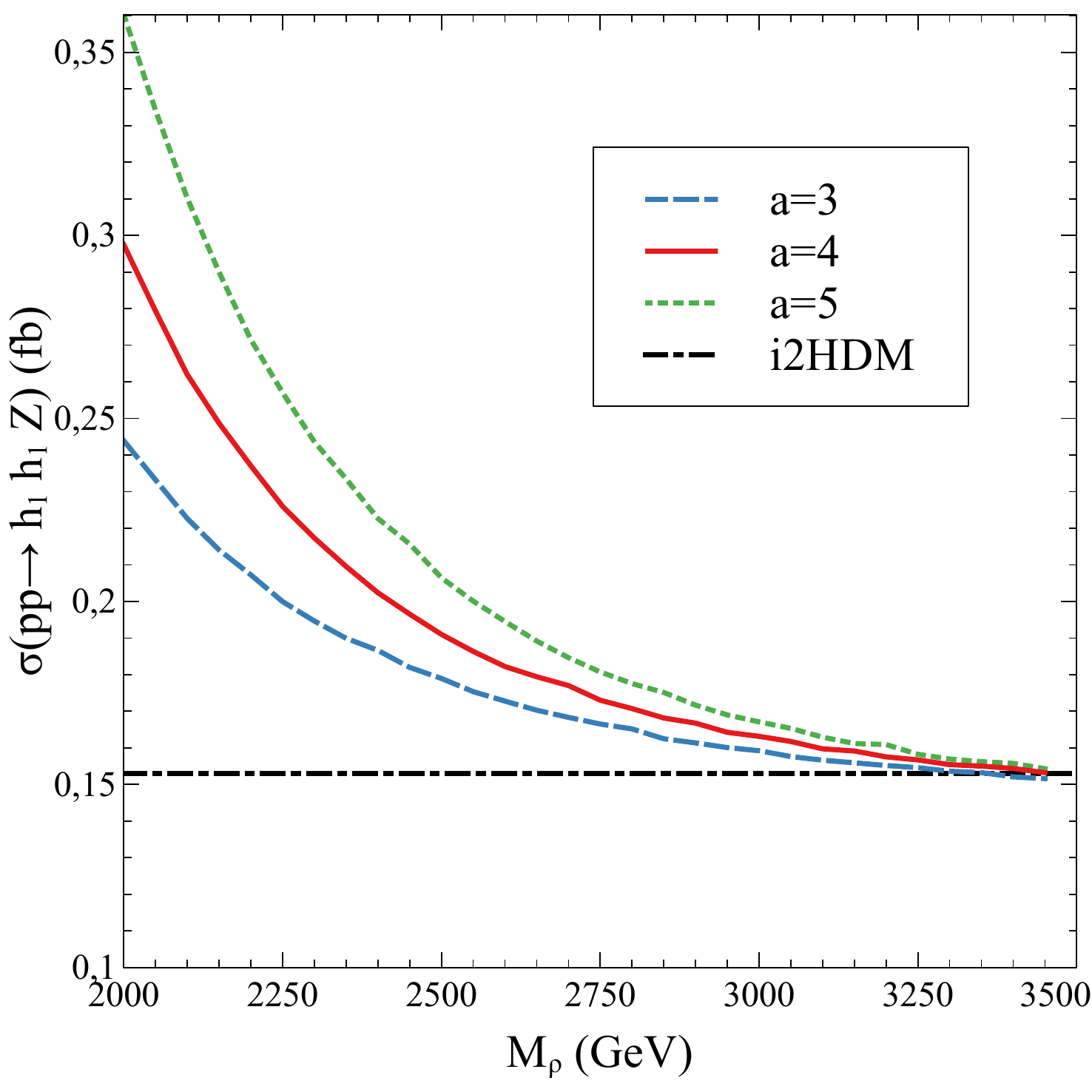}
	\includegraphics[width=0.45\textwidth]{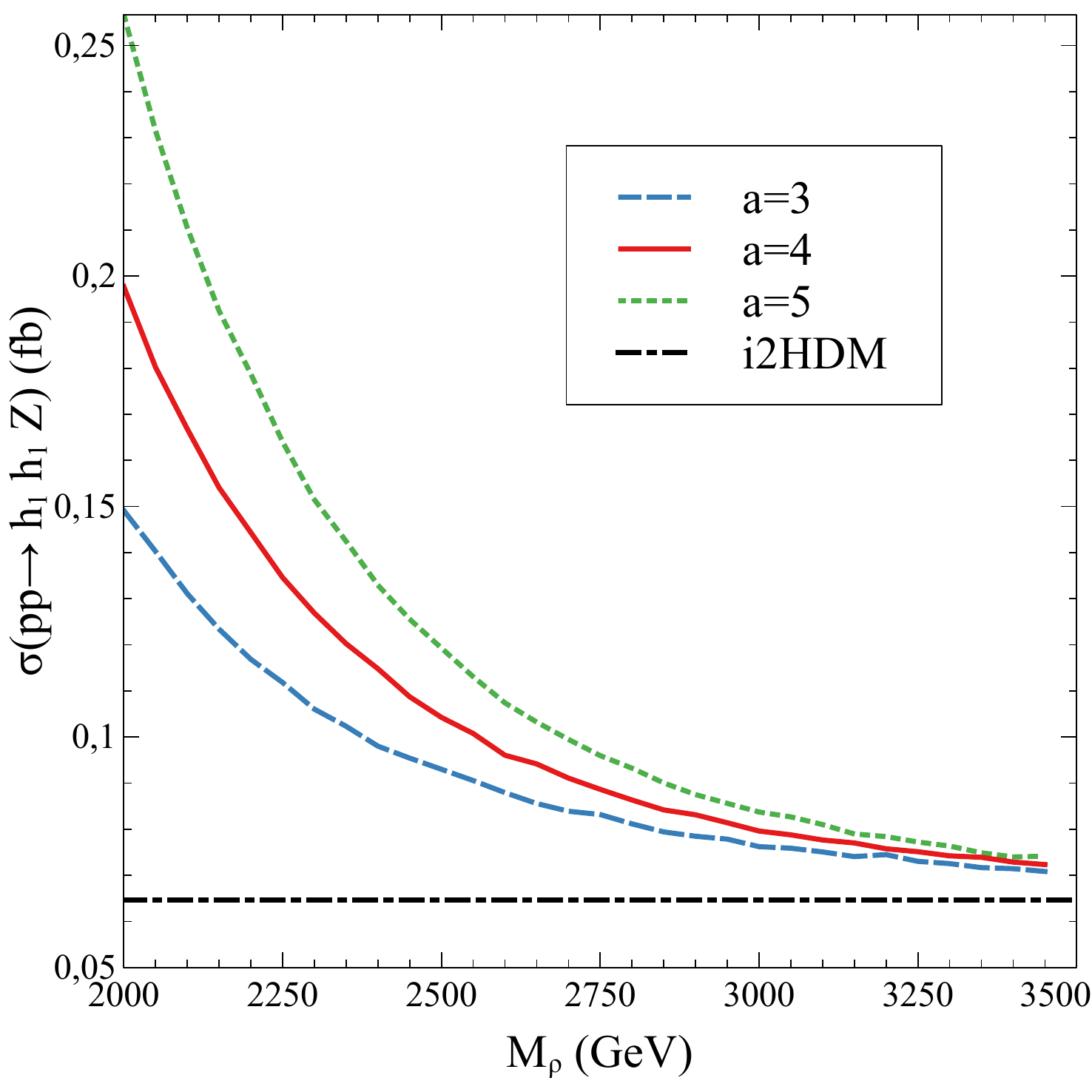}
	\caption{\textbf{Left}: $\sigma(pp\rightarrow h_1 h_1 Z)$  vs. $\Mrho$ at the $\sqrt{s}=13$ TeV LHC considering the benchmark point BM1. \textbf{Right}: \textit{Idem} but for the benchmark point BM6 }
	\label{fig:BM16}
\end{figure}

In order to compare the predictions of our model to the usual i2DHM ones, we compute $\sigma(pp\rightarrow h_1 h_1 Z)$ for the benchmark points 1 and 6 of reference \cite{Belyaev:2016lok} defined by $M_{h1}=55$ GeV, $M_{h2}=63$ GeV, $M_{h^{\pm}}=150$ GeV, $\ld[345]=1.0\times 10^{-4}$, $\ld[2]=1.0$ (BM1) and $M_{h1}=100$ GeV, $M_{h2}=105$ GeV, $M_{h^{\pm}}=200$ GeV, $\ld[345]=2.0\times 10^{-3}$, $\ld[2]=1.0$ (BM6) respectively. The  computed cross sections  include the kinematic cut $E_T \!\!\!\!\!\!\!/  \;\;> 100$ GeV for both benchmark points. In Figure \ref{fig:BM16}, we show our results, alongside the cross section predicted in the usual i2HDM, for BM1 (left) and BM6 (right). In both cases we can see an important enhancement in the low $\Mrho$ region compared to the usual i2DHM. 

\begin{figure}
	\includegraphics[scale=0.5]{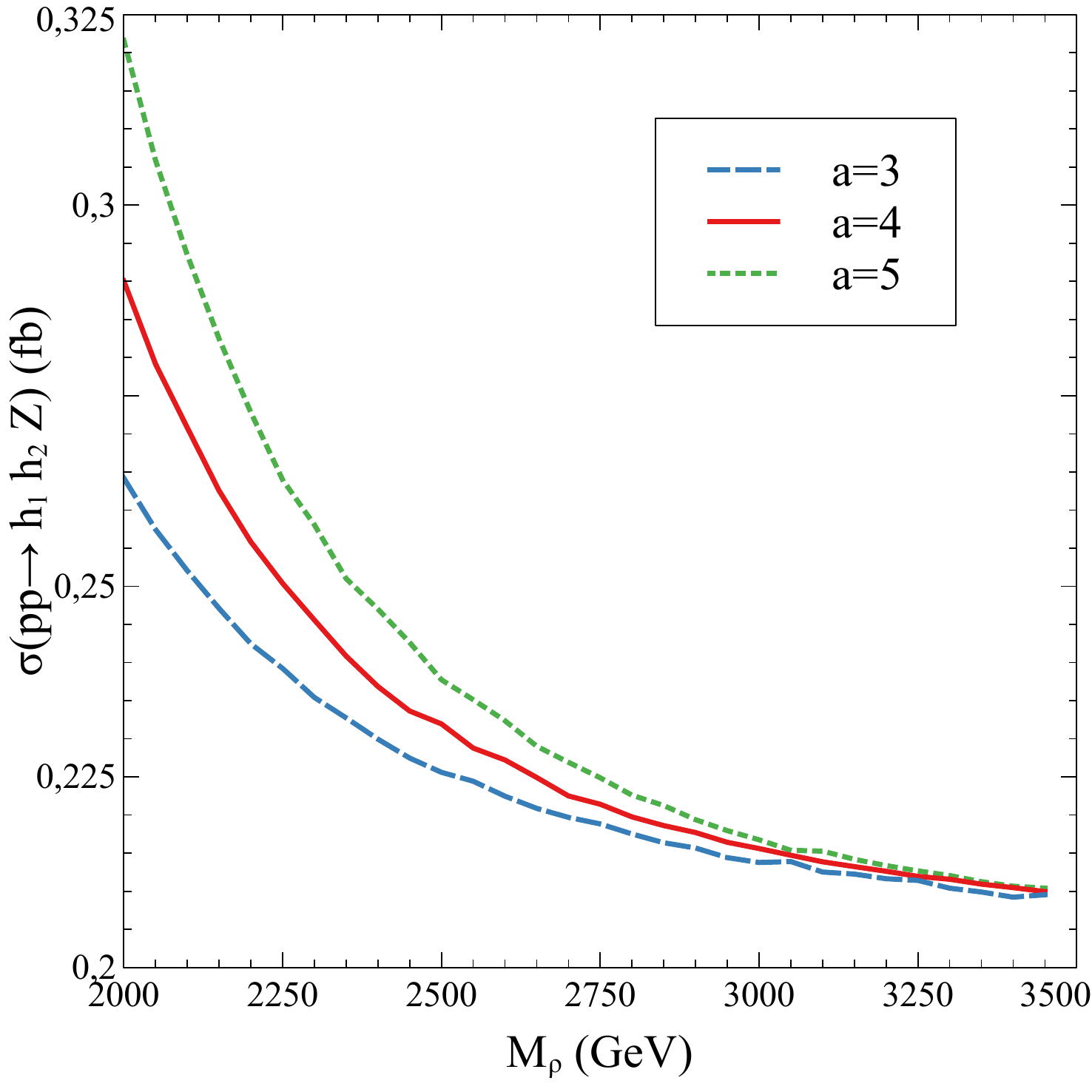}
	\caption{$\sigma(pp\rightarrow h_1 h_2 Z)$ vs. $\Mrho$ at the $\sqrt{s}=13$ TeV LHC considering the benchmark point BM6}
	\label{fig:h1h2Z}
\end{figure}

Additionally, we show in Figure \ref{fig:h1h2Z} our prediction for $\sigma(pp\rightarrow h_1 h_2 Z)$ at the $\sqrt{s}=13$ TeV LHC considering the benchmark point BM6. This process
also contributes to the mono-Z production provided that the mass splitting between $h_1$ and $h_2$ is small.

 \section{Conclusions}\label{Conclusions}
In this work, we have extended the i2DHM by adding a new heavy vector triplet and assuming that the inert scalar doublet is strongly coupled to the new spin-1 field. The theoretical construction was based on the Hidden Local Symmetry idea and thus the new vector field was introduced by enlarging the gauge symmetry to $SU(2)_{1}\times SU(2)_{2}\times U(1)_Y$. The hypothesis of a strong interaction between the heavy vector field and the inert scalar doublet was implemented making the inert scalar field to be a doublet of $SU(2)_{2}$ while the standard field (including the Higgs filed) were supposed to transform non-trivialy only under $SU(2)_{1}$.   

In general, the model is allowed by current data provided that $\Mrho > 2.4$ TeV but lower values of $\Mrho$ are possible when the decay of the new vector into non-standard scalar is open. Indeed, in this kinematic region the discovery of $\rho$ seems to be rather challenging at the LHC specially when it is considered its decay only into standard particles.  A more interesting possibility is the production of a $Z$ boson in association with two $h_1$ particles since the total process ($\rho$ production and decay) is less suppressed than the previous case. Naturally, the  $h_1$ particles would escape detection but they will produce a significant amount of missing transverse momentum. However, the predicted cross sections are quite small, although an important enhancement with respect to the usual i2DHM is observed for lower values of $\Mrho$, lying in the [0.1-0.3] fb range.

 However, the presence of the new heavy vector is not innocuous for the phenomenology of the Dark Matter candidate. In fact, it introduces new annihilation channels which are important in the region of large Dark Matter mass. The most important consequence of this phenomenon is the reduction of the relic density saturation zone compared with the usual i2DHM.


\section*{Acknowledgement}
This work was supported in part by Conicyt (Chile) grants ACT-146 and
PIA/Basal FB0821, and by  Fondecyt (Chile) grant 1160423.

\bibliographystyle{utphys}
\bibliography{bib}

\end{document}